\documentclass[aps,prstab,reprint,twocolumn,groupedaddress,showpacs,amsmath,amssymb]{revtex4-1}
\usepackage{epsfig}
\usepackage{amsmath}
\usepackage{dcolumn}
\usepackage{bm}
\usepackage{esint}
\usepackage{xcolor}

\newcommand{\degs}{$^{\circ}$}

\begin{document}


\title{Considerations and findings on beam vorticity dynamics}


\author{L.~Groening}
\affiliation{GSI Helmholtzzentrum f\"ur Schwerionenforschung GmbH, Darmstadt D-64291, Germany}


\date{\today}

\begin{abstract}
This document is on considerations and findings on modelling of spinning beams. Spinning has been proposed for stabilizing beams against perturbations notably risen by non-linear space charge forces, see~[Y.-L.~Cheon et al., Effects of beam spinning on the fourth-order particle resonance of 3D bunched beams in high-intensity linear accelerators, Phys. Rev. Accel. \& Beams {\bf 25}, 064002 (2022)]. Although not further treated therein, spinning can be quantified by angular momentum or by vorticity. Considering vorticity revealed that the latter has remarkable similarity w.r.t.~its modelling along solenoid channels to modelling the beam envelope. Matrices of vorticity transport, corresponding phase advances, and Twiss parameters look very similar and are partially even identical to their counterparts concerning envelopes. Corresponding to emittance, the quantity of vortissance, being a constant of motion, is defined. Unlike emittance, for vorticity-dominated beams it may take imaginary values, causing Twiss parameters, and negative or zero phase advances along a finite beam line section. This imposes considerable consequences on respective periodic solutions.
\end{abstract}


\maketitle


\section{Introduction}
\label{s_intro}
Preservation of beam quality is a major concern within the design of almost all linear particle accelerators. The quality can be degraded along the accelerator by various perturbations. The probably most known and feared perturbation are non-linear forces from the beam's self fields. Many techniques have been applied successfully in order to minimize the impact of space charge. Among these is provision of very regular, i.e., periodic focusing lattices as well as matching of the beam envelope to the latter. Matching is achieved if the beam's spatial parameters reproduce the periodicity of the lattice. Since decades matching has been restricted to the horizontal, vertical, and longitudinal bunched beam dimension, disregarding eventual coupling among the three planes of phase space (planes for brevity). Examples for this can be found for instance in~\cite{Sacherer,groening_prstab2008}. During the last years, extension of matching towards consideration of inter-plane coupling has been started~\cite{Khan_NIMa,hoover_prab2021,Chen_arxiv2023}. A very first conceptual approach has been sketched already in the 1980ies by~\cite{Chernin}.

Apart from matching, the amount of focusing strength has been optimized in order to minimize emittance growth along the lattice. To a large extend this implies avoiding single-particle resonances and collective instabilities, being reviewed in~\cite{Cheon_pop2020} for instance.

Recently, an additional tool for further reduction of remaining emittance growth has been proposed~\cite{Cheon_prab}. It is the controlled spinning of the beam, in analogy to stabilization of flying objects against turbulences. Some evidence for mitigation of emittance growth with increased spinning has been provided, hence paving path to a broad field of further research.

The reported activities herein aim for provision of tools for better understanding of what type of spinning or rotation stabilizes particle beams. Apart from angular momentum, rotation may be quantified by vorticity. For rigidly rotating objects with cylindrical symmetry, like balls, bullets or frisbees, the angular momentum is equal to the vorticity. However, particle beams are generally neither rigid nor cylindrically symmetric. Accordingly, the question rises, what type of rotation causes stabilization of beams.

Within the pursue of this question, the behaviour of vorticity has been investigated along solenoid channels, since the latter preserve angular momentum. It has been found that vorticity has some remarkable features leading to vorticity dynamics being very similar to envelope dynamics. The related findings shall be reported here.

The manuscript commences by mentioning some relevant features of angular momentum and vorticity related to particle beams. Afterwards, a beam line is sketched that can form beams having either angular momentum, or vorticity, or both. The fourth section is on properties of vorticity along linear lattice elements. The resulting vorticity beam dynamics modelling is described subsequently, followed by a section on properties of matched beams depending on their amount of vorticity. Finally, the properties w.r.t.~vorticity of a special pair of quadrupole triplets are briefly reported. The manuscript closes with preliminary conclusions and an outlook. Several well-known issues and equations from envelope dynamics are referred to throughout the manuscript. They have been placed into the appendices, in order to focus the main body of the report on the new topics.

\section{angular momentum, vorticity, and eigen-emittances}
\label{s_AngVorEig}
One issue not being addressed so far is the proper definition of spinning. For the time being, spinning or rotation has been associated and even been set equal to the amount of angular momentum. This seemed quite justified, considering the angular momentum's outstanding role in physics. In beam physics, the rms angular momentum can be properly defined through two second order beam moments as
\begin{equation}
	L\,:=\,\langle xy' \rangle - \langle x'y \rangle \,,
\end{equation}
where the brackets indicate the mean value of their content for a given ensemble, here the product of two particle coordinates. The above equation is the translation of~$\vec{r}\times\vec{p}$ into beam physics in transverse coordinates.

Throughout the manuscript, the horizontal particle position~$x$ points towards the left and~$y$ is upwards. The beam propagation~$s$ follows a right-handed coordinate system as $\vec{s}:=\vec{x}\times\vec{y}$. The notation~$u'$ indicates the derivative of the coordinate~$u$ w.r.t.~$s$.

Spinning or rotation may also be quantified by the vorticity for instance. It is defined by the local rotation of the velocity~$\vec{\nabla}\times\vec{v}$, being integrated over the extension of the ensemble.
\begin{equation}
	\tilde{\mathcal{V}}_A \,=\,\int\limits_A \,\left[\vec{\nabla}\times\vec{v}\right]\cdot d\vec{A}\,,
\end{equation}
Translated into beam physics and using~\cite{groening_prab2018}
\begin{equation}
\mathcal{V}\,=\,\,\langle y^2 \rangle \langle xy'\rangle - \langle x^2\rangle\langle yx'\rangle + \langle xy\rangle (\langle xx'\rangle - \langle yy'\rangle)
\end{equation}
results into
\begin{equation}
	\mathcal{V}_A\,:=\,\frac{\mathcal{V}}{A}\,,
\end{equation}
with $A^2:=\langle x^2\rangle \langle y^2\rangle - \langle xy\rangle ^2$ as the beam rms area. The quantity~$\mathcal{V}_A $ has the same dimension as angular momentum and projected transverse emittance.

\begin{figure}[hbt]
	\centering
	\includegraphics*[width=70mm,clip=]{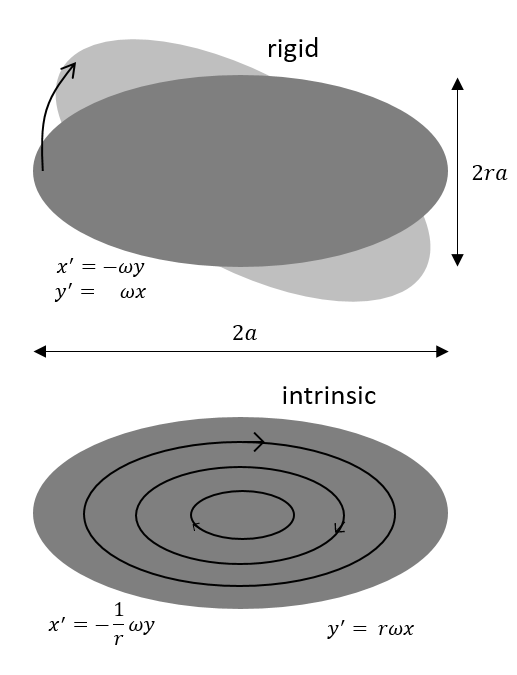}
	\caption{Ellipse with aspect ratio~$r$ performing a rigid rotation (upper) and an intrinsic rotation (lower).}
	\label{f_ellipses}
\end{figure}

Figure~\ref{f_ellipses} illustrates an example to distinguish angular momentum from vorticity. Particles forming ellipses perform rotations of different kind. The first rotation is rigid and all particles spin at angular velocity~$\omega$. The second scenario is an intrinsic rotation. Particles move around the centre without changing the ellipse shape.
Determining the angular momentum and $\mathcal{V}_A$ reveals for the rigid rotation~\cite{groening_prab2021}
\begin{equation}
	L_{rig}\,=\,\frac{\omega}{4}a^2(1+r^2)\,,
\end{equation}
\begin{equation}
	\mathcal{V}_{A,rig}\,=\,\frac{\omega}{2}a^2r\,,
\end{equation}
while for the intrinsic rotation one obtains
\begin{equation}
	L_{int}\,=\,\frac{\omega}{2}a^2r\,,
\end{equation}
\begin{equation}
	\mathcal{V}_{A,int}\,=\,\frac{\omega}{4}a^2(1+r^2)\,.
\end{equation}
The expressions for~$L$ and~$\mathcal{V}_A$ flip when flipping from rigid to intrinsic rotation. Additionally, for extreme aspect ratios of~$r\ll 1$ or~$r\gg 1$, the rigid rotation has just angular momentum and vanishing~$\mathcal{V}_A$ (relatively), while the intrinsic rotation has just~$\mathcal{V}_A$ but vanishing angular momentum (relatively). Another special case is the circle~($r=1$) with~$L$=$\mathcal{V}_A$. As shown in section~\ref{s_LaVBeams}, beams can be created having considerable angular momentum and zero vorticity and vice versa.

Vorticity received attention in beam physics thanks to its tight relation to eigen-emittances. The latter have been introduced in~\cite{Dragt} and are two constants of motion along symplectic beam line elements, which may couple the horizontal and vertical planes. The two eigen-emittances are equal to the projected rms emittances, once inter-plane coupling has been fully removed by symplectic elements. Their product is equal to the four-dimensional (4d) rms emittance.

The eigen-emittances of the two ellipses are calculated as (see App.~\ref{EigSymp} and~\cite{groening_prab2021})
\begin{equation}
	\epsilon_{1,rig}\,=\,\mathcal{V}_{A,rig}\,\leq\,L_{rig}\,,
\end{equation}
\begin{equation}
	\epsilon_{2,rig}\,=\,0
\end{equation}
and
\begin{equation}
	\epsilon_{1,int}\,=\,\mathcal{V}_{A,int}\,\geq\,L_{int} \,,
\end{equation}
\begin{equation}
	\epsilon_{2,int}\,=\,0\,.
\end{equation}
These relations are one example illustrating that eigen-emittances are related to vorticity rather than to angular momentum. Vorticity also occurs within the extension of Busch's theorem~\cite{Busch} to particle beams~\cite{groening_prab2018}. Finally, as shown in~\cite{groening_prab2021}, the change of eigen-emittances along short but non-symplectic beam line elements is related to the change of vorticity~$\mathcal{V}_A$ through 
\begin{equation}
	\label{e_square_diff_const}
	(\Delta\varepsilon )^2-\mathcal{V}_A^2\,=\,const\,,
\end{equation}
where $\Delta\varepsilon$ is the difference of the two eigen-emittances. As pointed out in~\cite{groening_prab2021}, the above equation is the generalization of Kim's relation~\cite{Kim}
\begin{equation}
	\label{e_Kim}
	\epsilon_{1/2}\,=\,\epsilon_{rms}\pm\frac{L}{2} 
\end{equation}
between eigen-emittances, projected rms-emittances, and angular momentum. Kim's relation applies to the special case of full cylindrical symmetry, i.e., $L$=$\mathcal{V}_A$ and with the constant of Eq.~(\ref{e_square_diff_const}) being equal to zero.

\section{Formation of L- and $\mathcal{V}_A$-beams}
\label{s_LaVBeams}
\begin{figure}[hbt]
	\centering
	\includegraphics*[width=75mm,clip=]{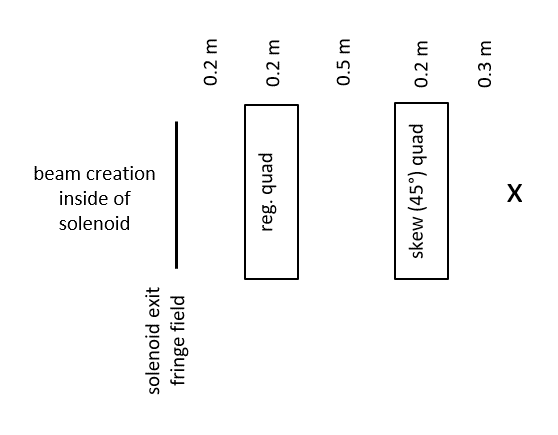}
	\caption{Beam line to form beams with arbitrary amounts of angular momentum~$L$ and vorticity$~\mathcal{V}_A$.}
	\label{f_SpinningLine}
\end{figure}

As pre-requisites shall be just re-called the statements of~\cite{groening_prab2021} concerning one property of regular and skewed quadrupoles. Both practically do not change the vorticity~$\mathcal{V}_A$. Figure~\ref{f_SpinningLine} depicts a beam line that can form beams with adjustable amounts of vorticity~$\mathcal{V}_A$ and angular momentum~$L$. The creation of ions is accomplished inside of a solenoid. At creation, the beam is fully uncoupled, i.e., all off-diagonal beam moments are zero. It shall be a dc-beam of protons with a sharp energy of 95~keV. The 4d-emittance is set to 200~mm$^\text{2}$mrad$^\text{2}$, accordingly both eigen-emittances are equal to~14.1~mm~mrad. The beam rms width is chosen to be 5.3~mm in both transverse planes. Extraction of the beam is through the exit fringe field of the solenoid. The exit fringe field imposes rotation to the cylindrical symmetric beam with~\cite{groening_prab2021}
\begin{equation}
	L\,=\,\mathcal{V}_A\,=\,-2\kappa A
\end{equation}
and $\kappa $:=$B/(2(B\rho))$, with $B$ as solenoid magnet field strength and $B\rho $ as beam rigidity.

Afterwards, a regular quadrupole removes the cylindrical symmetry. The quadrupole does practically not change~$L$ since the beam features $\langle xy \rangle $=0 at the quadrupole entrance ($L$ is not changed strictly just in the thin lens approximation). As mentioned above, it does not change the vorticity either. The regular quadrupole is followed by a skewed quadrupole. The beam at its entrance has no cylindrical symmetry, hence this quadrupole changes~$L$ (but not~$\mathcal{V}_A$). For a short skewed quadrupole, $\Delta L$ is calculated as (using Eq.~(\ref{e_m_singpart_skewquad}))
\begin{equation}
	\Delta L\,=\,k\cdot ( \langle y^2\rangle - \langle x^2\rangle)\,,
\end{equation}
where $k$:=$GL_{qs}/(B\rho)$ with $G$ as the field gradient and~$L_{qs}$ as the effective field length.

Accordingly, the sketched beam line allows for imposing arbitrary amounts of~$L$ and~$\mathcal{V}_A$ to the beam. In summary, the method is to impose the required~$\mathcal{V}_A$ through the solenoid field strength and to adapt the quadrupole strengths to the desired~$L$. Although the method has been described assuming short quadrupoles, the principle works also in general with quadrupoles of finite lengths.

\begin{figure}[hbt]
	\centering
	\includegraphics*[width=88mm,clip=]{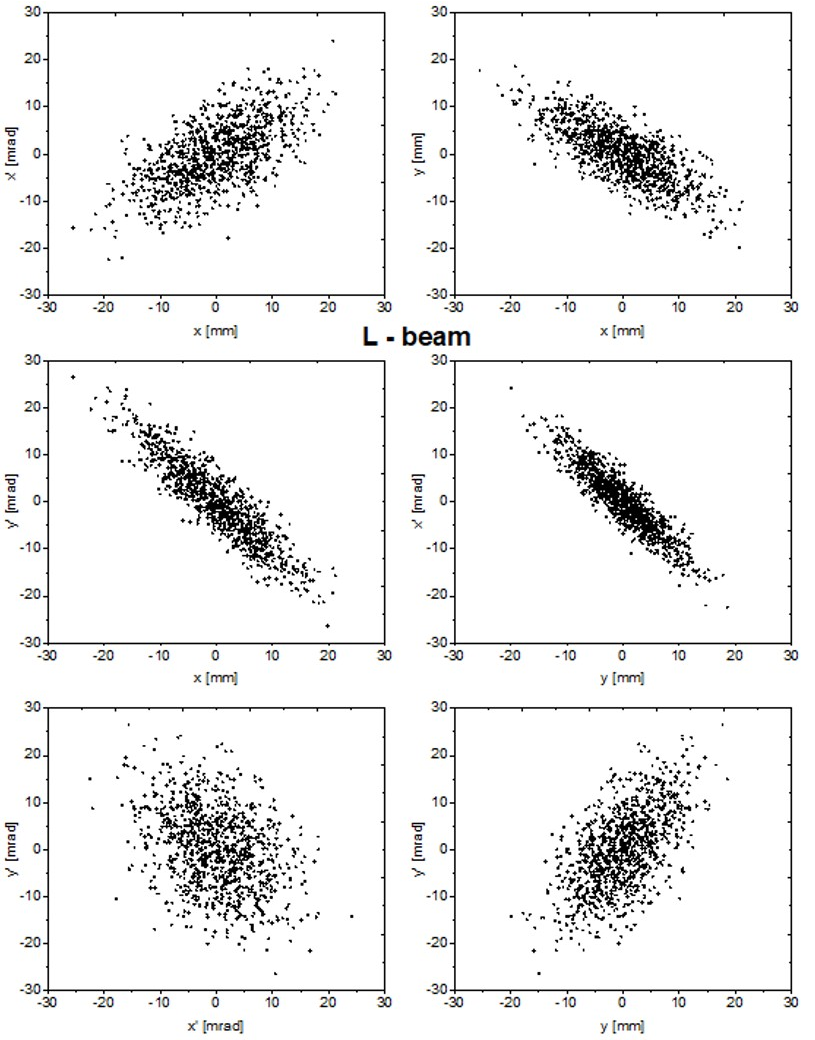}
	\caption{Two-dimensional projections of a~$L$-beam, $\varepsilon _1$=$\varepsilon _2$=14~mm~mrad, $L$=-20~mm~mrad, $\mathcal{V}_A$=0, $\varepsilon _x$=$\varepsilon _y$=46~mm~mrad.}
	\label{f_LBeam}
\end{figure}

\begin{figure}[hbt]
	\centering
	\includegraphics*[width=88mm,clip=]{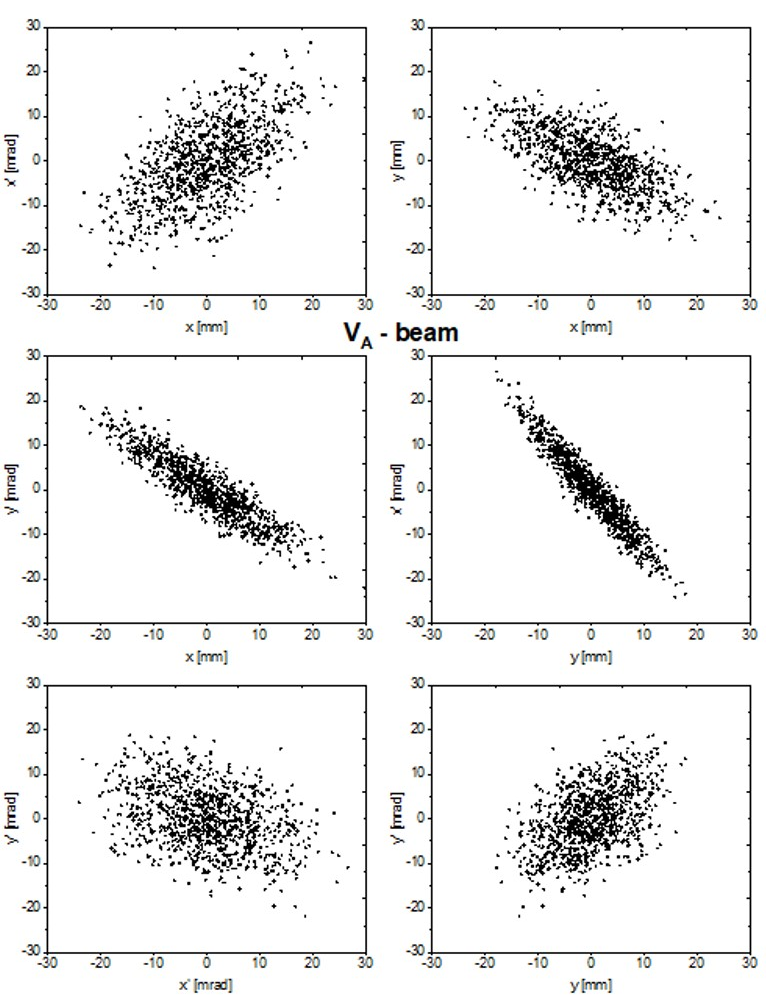}
	\caption{Two-dimensional projections of a~$\mathcal{V}_A$-beam. $\varepsilon _1$=27~mm~mrad, $\varepsilon _2$=7~mm~mrad, $L$=0, $\mathcal{V}_A$=20~mm~mrad, $\varepsilon _x$=59~mm~mrad, $\varepsilon _y$=40~mm~mrad.}
	\label{f_VBeam}
\end{figure}

\begin{figure}[hbt]
	\centering
	\includegraphics*[width=88mm,clip=]{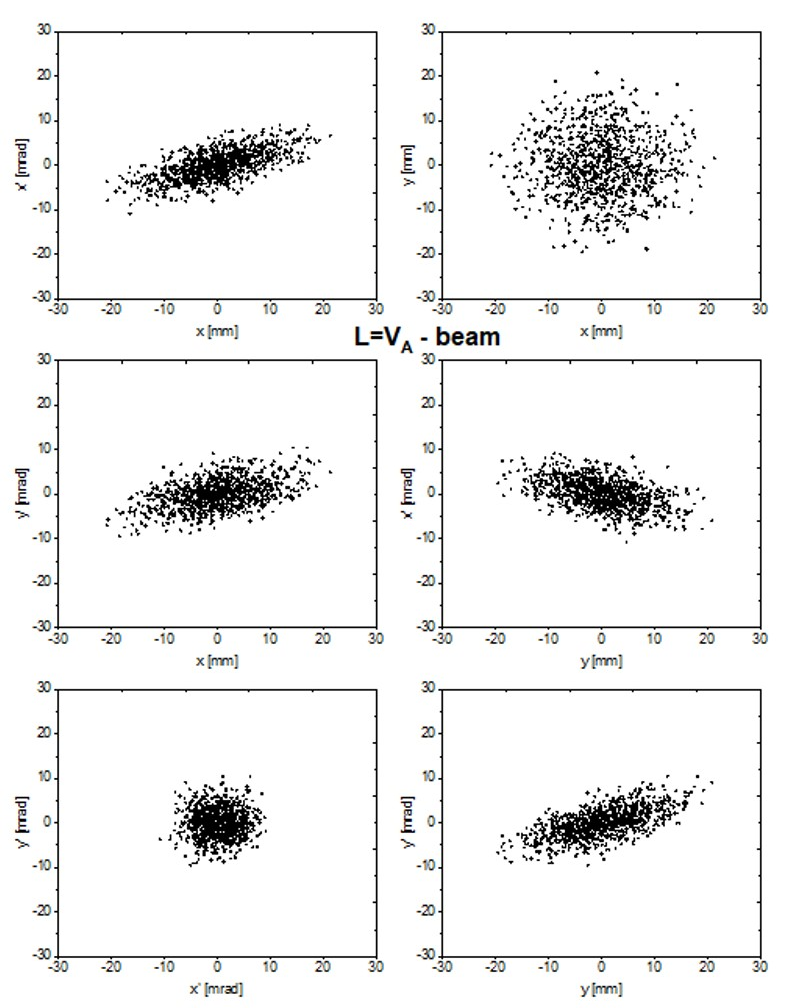}
	\caption{Two-dimensional projections of a~$L$=$\mathcal{V}_A$-beam. $\varepsilon _1$=27~mm~mrad, $\varepsilon _2$=7~mm~mrad, $L$=$\mathcal{V}_A$=20~mm~mrad, $\varepsilon _x$=$\varepsilon _y$=17~mm~mrad.}
	\label{f_LVBeam}
\end{figure}

Table~\ref{tab_beamline} lists the settings of the beam line which form various amounts of angular momentum and vorticity~$\mathcal{V}_A$. The resulting phase space distribution for a beam with angular momentum but without~$\mathcal{V}_A$ is plotted in~Fig.~\ref{f_LBeam}. Such a beam is referred to as an $L$-beam. It has equal eigen-emittances in agreement to Eq.~(\ref{e_square_diff_const}) with the constant being equal to zero.  Figure~\ref{f_VBeam} instead depicts the distribution of a beam without angular momentum but with~$\mathcal{V}_A$, which will be referred to as a~$\mathcal{V}_A$-beam. This beam has different eigen-emittances, again in agreement to Eq.~(\ref{e_square_diff_const}). For completeness, the distribution for a beam with angular momentum equal to vorticity~$\mathcal{V}_A$ is shown in~Fig.~\ref{f_LVBeam}. This beam has cylindrical symmetry and is referred to as an $L$=$\mathcal{V}_A$-beam.
\begin{table}[hbt]
	\caption{Magnet settings for the beam line to form various amounts of angular momentum~$L$ and vorticity~$\mathcal{V}_A$.}
	\begin{tabular}{c|c|c|c}
		$L/\mathcal{V}_A$ & solenoid field & reg. quad. grad. & sk. quad. grad. \\
		$\mu$m/$\mu$m & mT & mT/m & mT/m \\
		\hline
        0.0 / 0.0 &   0 &   0 &    0 \\
        \hline
       -20 / 0.0 &   0 & 417 & -280 \\
        0.0 / 20 & 321 & 417 & -273 \\
        20 / 20  & 314 &   0 &    0 \\
        \hline
       -40 / 0.0  &   0 & 417 & -565 \\
        0.0 / 40  & 664 & 417 & -479 \\
        40 / 40   & 630 &   0 &    0 \\
        \hline
	\end{tabular}
	\label{tab_beamline} 
\end{table}
It shall be mentioned that the beam parameters listed within the captions of Figs.~\ref{f_LBeam}, \ref{f_VBeam}, and \ref{f_LVBeam}, all meet the general Eq.~(\ref{e_square_diff_const}), while just the special case of cylindrical symmetry (Fig.~\ref{f_LVBeam}) meets~Eq.~(\ref{e_Kim}).

\section{Properties of beam vorticity}
The beam vorticity and its derivatives w.r.t.~$s$ along a drift are
\begin{equation}
	\mathcal{V}\,=\,\,\langle y^2 \rangle \langle xy'\rangle - \langle x^2\rangle\langle yx'\rangle + \langle xy\rangle (\langle xx'\rangle - \langle yy'\rangle)\,,
\end{equation}
\begin{equation}
	\begin{split}
		\mathcal{V}'\,=\,&\left[\langle xx' \rangle + \langle yy' \rangle\right] \left[\langle xy' \rangle - \langle x'y \rangle\right]\\
		&+\,  \langle x'y' \rangle\left[\langle y^2 \rangle - \langle x^2 \rangle\right]\\
		&+\,  \langle xy \rangle\left[\langle x'^2 \rangle - \langle y'^2 \rangle\right]\,,
	\end{split}
\end{equation}
\begin{equation}
	\begin{split}
		\mathcal{V}''\,=\,&2\left[\langle x'^2 \rangle\langle xy'\rangle - \langle y'^2 \rangle\langle x'y \rangle\right]\\
		+\,&2\langle x'y' \rangle\left[\langle yy' \rangle - \langle xx' \rangle\right]\,,
	\end{split}
\end{equation}
and
\begin{equation}	
	\mathcal{V}'''\,=\,0\,.	
\end{equation}
As for the beam envelope~$\langle x^2 \rangle $, the second derivative of~$\mathcal{V}$ is a constant along a drift. Accordingly, vorticity is transformed through a drift of length~$d$ by the matrix equation
\begin{equation}
	\label{e_m_drift_eta_3}
	\begin{bmatrix}
		\mathcal{V} \\
		\mathcal{V}' \\
		\mathcal{V}''
	\end{bmatrix}\,\rightarrow\,
	\begin{bmatrix}
		1 & d & \frac{d^2}{2} \\
		0 & 1 & d \\
		0 & 0 & 1
	\end{bmatrix}\cdot
	\begin{bmatrix}
		\mathcal{V} \\
		\mathcal{V}' \\
		\mathcal{V}''
	\end{bmatrix}\,,
\end{equation}
which is in full analogy to the transformation of the envelope by a drift given in~Eq.~(\ref{e_m_drift_e_3}). Equation~(\ref{e_m_drift_eta_3}) introduces the vorticity vector~$\vec{\mathcal{V}}$.

Angular momentum is strictly preserved along solenoids, while vorticity is changed. First, the transport of vorticity along a short solenoid shall be determined. This starts from the transformation of beam moments by a short solenoid by the matrix of Eq.~(\ref{e_m_singpart_solshort}). Through re-grouping of the initial and final beam moments, the resulting final vorticity vector~$\mathcal{V}$ can be expressed through the initial vorticity vector. The transformation can be stated as
\begin{equation}
	\label{e_m_solshort_eta_3}
\vec{\mathcal{V}}\,\rightarrow\,
	\begin{bmatrix}
		1 & 0 & 0 \\
		-2\kappa ^2L_s & 1 & 0 \\
		2\kappa ^4 L_s^2 & -2\kappa ^2L_s & 1
	\end{bmatrix}\cdot
\vec{\mathcal{V}}\,.
\end{equation}
Doing so reveals, that the transformation of the beam envelope through a short quadrupole by Eq.~(\ref{e_m_shortquad_e_3}) and the transformation of~$\mathcal{V}$ through a short solenoid are identical if~$k=\kappa ^2 L_s$.

The same method is applied to derive the transformation of vorticity by a solenoid of arbitrary length. Equation~(\ref{e_m_singpart_sol}) delivers the transformation of beam moments through a general solenoid. As for the short solenoid, re-grouping of the final beam moments results into the desired transformation of~$\mathcal{V}$. This re-grouping is very lengthy and it takes careful book keeping of the numerous terms. The final result is the transformation of vorticity as
\begin{equation}
	\label{e_m_sol_eta_3}
	\vec{\mathcal{V}}\,\rightarrow\,
	\begin{bmatrix}
		C^2 & CS/\kappa & S^2/(2\kappa ^2) \\
		-2\kappa CS & C^2-S^2 & CS/\kappa \\
		2\kappa ^2 S^2 & -2\kappa CS & C^2
	\end{bmatrix}\cdot
	\vec{\mathcal{V}}\,.
\end{equation}
The above equation is the counterpart of Eq.~(\ref{e_m_quad_e_3}) which transports the beam envelope through a finite regular and focusing quadrupole.

To summarize, the transformation of vorticity along drifts and solenoids is modelled by matrix equations. The involved matrices have the determinant of~1 and have counterparts within the transformation of the beam envelope.

Additionally, the transformation of vorticity through instantaneous angular kicks as
\begin{equation}
	x'\,\rightarrow\,x'-k_xx
\end{equation}
\begin{equation}
	y'\,\rightarrow\,y'-k_yy
\end{equation}
shall be reported. If $k_x$=$k_y$=$k>0$, this corresponds to a short solenoid with $\kappa ^2$=$(k/L_s)$. If both are different from each other but positive, it models space charge kicks for an elliptical but non-round beam with homogeneous spatial distribution. The vorticity vector is changed as
\begin{equation}
	\mathcal{V}\,\rightarrow\,\mathcal{V}
\end{equation}
\begin{equation}
	\mathcal{V}'\,\rightarrow\,\mathcal{V}'\,-\,\left[k_x+k_y\right]\mathcal{V}
\end{equation}
\begin{equation}
	\label{e_angkicks_eta}
	\begin{split}
		\mathcal{V}''\,\rightarrow\,\mathcal{V}''\,& + 2k_x\cdot\left[ \langle x^2 \rangle \langle x'y' \rangle + \langle xy \rangle \langle y'^2 \rangle \right] \\
		& - 2k_x \cdot \langle xy' \rangle \cdot \left[\langle xx' \rangle + \langle yy' \rangle \right] \\
		& - 2k_y \cdot \left[ \langle y^2 \rangle \langle x'y' \rangle + \langle xy \rangle \langle x'^2 \rangle \right] \\
		& + 2k_y \cdot \langle x'y \rangle \cdot \left[\langle xx' \rangle + \langle yy' \rangle \right] \\
		& + 2k_xk_y\cdot \mathcal{V}\,.
	\end{split}
\end{equation}
In general, this transformation cannot be expressed through a matrix operation on~$\vec{\mathcal{V}}$. It can be done just if~$k_x$=$k_y$=$k$, i.e., if the~$k$-terms in~$\mathcal{V}''$ sum up to~-2$k\mathcal{V}'$. Accordingly, if~$k_x\neq k_y$, the effect of linear angular kicks and hence of quadrupoles on~$\vec{\mathcal{V}}$ cannot be modelled by a matrix.

This section closes by considering the transformation of the vorticity vector through a solenoid exit fringe field. For the time being, no general expressions have been found, which are reasonably compact. For the simple case of an initially fully uncoupled beam, i.e., of zero off-diagonal moments, the change is given by
\begin{equation}
	\label{e_transp_solfringe_v}
	\mathcal{V}\,\rightarrow \,-2\kappa A^2\,,
\end{equation}
\begin{equation}
	\label{e_transp_solfringe_vp}
	\mathcal{V}'\,\rightarrow \,0\,,
\end{equation}
\begin{equation}
	\label{e_transp_solfringe_vpp}
	\mathcal{V}''\,\rightarrow \,-2\kappa\,\left[\varepsilon _1 ^2 + \varepsilon _2 ^2 +2\kappa ^2 A^2 \right]\,,
\end{equation}
which cannot be expressed by a matrix equation.

\section{Vorticity beam dynamics and vortissance}
This section introduces the term of beam vortissance, which is a quantity being preserved along drifts and solenoids. Corresponding transport matrices are defined as well as phase advances and Twiss parameters. Special emphasis is put on the fact and consequences that vortissance can be purely imaginary.

\subsection{Beam vortissance}
\label{ss_vortissance}
The paragraph commences by realizing the similarity of Eq.~(\ref{e_m_drift_e_3}) to Eq.~(\ref{e_m_drift_eta_3}), Eq.~(\ref{e_m_shortquad_e_3}) to Eq.~(\ref{e_m_solshort_eta_3}), and Eq.~(\ref{e_m_quad_e_3}) to Eq.~(\ref{e_m_sol_eta_3}). These similarities suggest construction of a quantity from~$\vec{\mathcal{V}}$ which is preserved along drifts and by solenoids. To this end, the horizontal rms emittance shall be re-expressed as
\begin{equation}
	\epsilon_x \,=\,\sqrt{\langle x^2 \rangle \frac{\langle x^2 \rangle ''}{2} - \frac{\langle x^2 \rangle '^2}{4}} \,.
\end{equation}
It is preserved along drifts and regular quadrupoles. Accordingly, the "rms vortissance" shall be defined in analogue way as
\begin{equation}
	\label{e_V_const}
	V\,:=\,\sqrt{\mathcal{V} \frac{\mathcal{V} ''}{2} - \frac{\mathcal{V} '^2}{4}}\,.
\end{equation}
Using Eq.~(\ref{e_m_drift_eta_3}), Eq.~(\ref{e_m_solshort_eta_3}), and Eq.~(\ref{e_m_sol_eta_3}) confirms preservation of~$V$ by each of these transformations. It shall be explicitly mentioned that $V^2$ can be negative, hence the vortissance can take purely imaginary values. The unit of~$V$ is mm$^\text{2}$mrad$^\text{2}$. Since it is not positive definite, $V$ is neither equal to the 4d-rms emittance nor to the square of the angular momentum. In fact, even a beam without angular momentum can feature a considerable amount of vortissance~$V$ as shown in section~\ref{s_LaVBeams}. The vortissance is not preserved along quadrupoles as indicated by Eq.~(\ref{e_angkicks_eta}) with $k_y$=$-k_x$ since then the latter is not a matrix equation which involves just derivatives of~$\mathcal{V}$.

For the time being, a reasonably short expression of~$V^2$ through second beam moments has not been found. However,~$V^2$ can be expressed as the determinant of the matrix
\begin{equation}
	\label{e_m_MV}
	W_v\,:=\,
	\begin{bmatrix}
		\mathcal{V} & \mathcal{V} '/2 \\
		\mathcal{V}'/2 & \mathcal{V} ''/2
	\end{bmatrix}\,,
\end{equation}
in straight analogy to the relation of beam moments matrix to emittance for the beam envelope.

\subsection{Vorticity transport matrices}
\label{ss_Vmatrices}
Exploiting the analogies derived above, transport matrices~$M_e$ are defined such, that for a given beam line element the transport of~$W_v$ is mathematically identical to the transport of the beam envelope, i.e.,
\begin{equation}
	\label{e_m_generell_eta}
	W_{v,f}\,:=\,M_e \cdot W_{v,i} \cdot M_e^{T}\,,\,\,\,\,\,det(M_e)=\text{1}\,.
\end{equation}

Using Eqs.~(\ref{e_m_drift_eta_3}), (\ref{e_m_solshort_eta_3}), and (\ref{e_m_sol_eta_3}) together with claiming the determinant of each~$M_e$ to be equal to~1, these matrices can be derived in a straight forward way. For a drift one obtains
\begin{equation}
	\label{e_m_drift_eta}
	M_{drift}\,=\,
\begin{bmatrix}
	1 & d \\
	0 & 1
\end{bmatrix}
\end{equation}
and for a short solenoid
\begin{equation}
	\label{e_m_shortsol_eta}
	M_{sol,short}\,=\,
	\begin{bmatrix}
		1 & 0 \\
		-\kappa ^2L_s & 1
	\end{bmatrix}\,,
\end{equation}
while for a general solenoid it is
\begin{equation}
	\label{e_m_sol_eta}
	M_{sol}\,=\,
	\begin{bmatrix}
		C & S/\kappa \\
		-\kappa S  & C
	\end{bmatrix}\,.
\end{equation}

Finally, the matrix~$M_{sds}$ corresponding to the single particle transport matrix~$m_{sds}$ (Eq.~(\ref{e_m_singpart_sds})) is simply constructed through
\begin{equation}
	M_{sds}(\kappa , d)\,:=\,M_{sol}(-\kappa)\cdot M_{drift}(d)\cdot M_{sol}(\kappa)
\end{equation}
and is found to read as
\begin{equation}
	M_{sds}(\kappa , d)\,=\,
	\begin{bmatrix}
		C^2 - \kappa dCS - S^2 &  2\frac{CS}{\kappa} + dC^2 \\
		-2\kappa CS + \kappa ^2 dS^2 &  C^2 - \kappa dCS - S^2
	\end{bmatrix}\,.
\end{equation}

There is identity of~$m_{drift,xx}$ and~$M_{drift}$, as well as of~$m_{sol,short,xx}$ and~$M_{sol,short}$ as well as of~$m_{sds,xx}$ and~$M_{sds}$. These identities imply very convenient consequences for lattices comprising drifts and pairs of solenoids with opposite field directions. For instance, the periodic phase advances of~$\langle x^2 \rangle (s)$ and of~$\mathcal{V}(s)$ along such lattices are identical, i.e.,
\begin{equation}
	\begin{split}
     cos(\Delta \Phi _{m,xx,latt})\,&=\,\frac{1}{2}Tr(m_{xx,latt}) \\
     &=\,\frac{1}{2}Tr(M_{latt})\,=\,cos(\Delta \Phi _{M,latt})\,.
     \end{split}
\end{equation}
Modelling the transport of the beam envelope~$\langle x^2 \rangle $ and of the vorticity~$\mathcal{V}$ is through the same matrices. As seen throughout the manuscript, the corresponding periodic Twiss parameters are equal as well. Accordingly, periodic solutions of~$\mathcal{V}$ within a sequence of solenoids and drifts can be constructed or determined using the same matrices as for periodic solutions for the beam envelope within a sequence of focusing quadrupoles and drifts.

\subsection{Vortissance Twiss parameters}
The vortissance~$V$ relates to the vorticity as the emittance relates to the beam envelope. However, since~$V^2$ may be negative, $V$ can take purely imaginary values. This issue is discussed in the next subsection. The case of~$V=0$ is not considered, since if it is equal to zero, there is no need to consider vorticity dynamics.

Thanks to the above correspondences, vorticity Twiss parameters can be defined as for the beam envelope through
\begin{equation}
	\beta _v\,:=\,\frac{\mathcal{V}}{V}\,,
\end{equation}
\begin{equation}
	\alpha _v\,:=\,-\frac{\beta _v '}{2}\,=\,\frac{-\mathcal{V}'}{2V}\,,
\end{equation}
\begin{equation}
	\gamma _v\,:=\,\frac{1+\alpha _v ^2}{\beta _v}\,=\,\frac{\mathcal{V}''}{2V}.
\end{equation}
The units of the vorticity Twiss parameters are the same as of the beam envelope Twiss parameters, i.e, m, 1, and~1/m. 

Vorticity betatron phase advances between two locations~$a$ and~$b$ along~$s$ may be defined as for the envelope through
\begin{equation}
	\label{e_PhaseAdvance_int}
	\Delta \Phi _{v,a,b}\,:=\,\int _a^b \frac{ds}{\beta _v (s)}\,,
\end{equation}
which along a drift turns into
\begin{equation}
	\Delta \Phi _{v,a,b}\,:=\,\pm \left[atan\left[\frac{\mathcal{V}''(s) s+\mathcal{V}'(s)}{2V}\right]\right]_a^b\,.
\end{equation}
Unlike the beam envelope and squared emittance, the vorticity and squared vortissance can take negative values. This has some consequences, which shall be discussed in the following.

\subsubsection{Real vortissance $V$}
\label{ss_Vreal}
If $V^2$ is positive, the vortissance is real, resulting into real Twiss parameters and phase advances. The determinant of the corresponding vorticity Twiss parameter matrix
\begin{equation}
	\begin{bmatrix}
		\beta _v & -\alpha _v \\
		-\alpha _v & \gamma _v
	\end{bmatrix}
\end{equation}
is equal to~1. However, the vorticity~$\mathcal{V}$ and its second derivative~$\mathcal{V}''$ may be negative. In consequence, the vorticity Twiss parameters~$\beta _v$ and~$\gamma _v$ will be negative as well as the phase advances. This does not occur for beam envelopes, which are positive by construction. Apart from this ambiguity in sign, there is full equivalence of vorticity dynamics and envelope dynamics along channels made from solenoids pairs of opposite field directions.

For such channels, there exist periodic solutions of the Twiss parameters with corresponding phase advances for both, the beam envelope and for the vorticity. One set of solutions is equal to each other,
\begin{equation}
	\Delta\Phi _{v,per} \,=\, \Delta\Phi _{e,per}\,>\,0
\end{equation}
\begin{equation}
	\beta _{v,per} \,=\, \beta _{e,per}\,>\,0
\end{equation}
\begin{equation}
	\alpha _{v,per} \,=\, \alpha _{e,per}\,.
\end{equation}
Since vortissance beam dynamics permits also for negative~$\beta _v$ and~$\gamma _v$, there is the second periodic vorticity solution with
\begin{equation}
	\Delta\Phi _{v,per} \,=\, -\Delta\Phi _{e,per}\,<\,0
\end{equation}
\begin{equation}
	\beta _{v,per} \,=\, -\beta _{e,per}\,<\,0
\end{equation}
\begin{equation}
	\alpha _{v,per} \,=\, -\alpha _{e,per}\,.
\end{equation}
Physically, the second solution just describes a beam with vortex of same amount but into the opposite direction w.r.t.~the first solution. Inverting the vortex direction will change the sign of~$\vec{\mathcal{V}}$ and hence preserve the vortissance~$V$.

The constance of~$V^2>0$ enforces both, $\mathcal{V}$ and~$\mathcal{V}''$, to be of same sign and being different from zero. Hence, $V^2>0$ imposes an intrinsic defocusing of $\mathcal{V}$ and~$\mathcal{V}''$ away from zero. The sign of both is preserved, i.e., the zeros cannot be crossed. In envelope dynamics, this is the well known emittance defocusing term.

\subsubsection{Imaginary vortissance $V$}
\label{ss_Vimaginary}
In case of~$V^2<0$, the Twiss parameters are purely imaginary and shall be defined as:
\begin{equation}
	\beta _v\,:=-i\,\frac{\mathcal{V}}{|V|}\,,
\end{equation}
\begin{equation}
	\alpha _v\,:=\,-\frac{\beta _v '}{2}\,=\,i\frac{\mathcal{V}'}{2|V|}\,,
\end{equation}
\begin{equation}
	\gamma _v\,:=\,\frac{1+\alpha _v ^2}{\beta _v}\,=\,-i\frac{\mathcal{V}''}{2|V|}.
\end{equation}
The determinant of the corresponding vorticity Twiss parameter matrix is equal to~-1. As for real vortissance, the Twiss parameters can assume negative (but imaginary) values. Phase advances are purely imaginary and can take negative values as well. In contrast to real vortissance, there is no intrinsic defocusing of vorticity. Accordingly, $\mathcal{V}$ and/or $\mathcal{V}''$  as well as the Twiss parameters $\beta _v$ and/or $\gamma _v$ may be zero, or different in sign.

However, $\mathcal{V}$=$\beta _v$=0 does not cause an ill-defined~$\gamma _v$, since the constance of~$V^2<0$ imposes~$\mathcal{V}'$=$2iV$ and hence \mbox{$\alpha _v ^2$=-1}, thus preventing the singularity of~$\gamma _v$. Additionally, the singularity is physically prevented by~$\gamma _v$=$\mathcal{V}''/2V$ with~$V\neq\,0$.

Eventual zero crossings of~$\beta _v$ do not harm determination of the imaginary phase advance according to Eq.~(\ref{e_PhaseAdvance_int}). But they may result into zero phase advance between two locations. This does not occur in envelope dynamics nor in vorticity dynamics at real vortissance.

The possibility of zero crossings of vorticity and of vanishing phase advances has impact on the nature of periodic solutions. For instance, the phase advance along a periodic cell is given by the trace of the matrix modelling this cell. The traces of the vortissance matrices~$M_e$ are positive and real each. But anyway, at imaginary vortissance, phase advances are imaginary. Accordingly, the definition of phase advance through the trace of the periodic cell matrix shall be extended, such that for imaginary vortissances, it must be multiplied by~$\pm i$ as
\begin{equation}
	\label{e_PhaseAdvanceTrace_imaginary}
	cos(\Delta \Phi _{v,per})\,=\,\pm i\frac{1}{2}Tr(M_e)\,,\,\,\,\,\,\text{if}\,\,\,V^2<0\,.
\end{equation}

Traces of periodic cell matrices are always positive. This applies to periodic envelope cells as well as to periodic vorticity cells. Hence, periodic solutions with zero phase advances cannot exist neither in envelope nor in vorticity dynamics. This applies even for the extended definition of periodic vorticity phase advance through~Eq.~(\ref{e_PhaseAdvanceTrace_imaginary}). Hence, there are no periodic vorticity solutions that represent symmetric quasi-oscillations around zero of the kind
\begin{equation}
	\beta _v(s+L_c/2)\,=\,-\beta _v(s)\,,
\end{equation}
with $L_c$ as the length of one cell.

\subsection{Sources of vortissance}
Transformation of vorticity by quadrupoles and by solenoid fringe fields cannot be expressed by symplectic matrices. In fact, these two beam line elements are sources (or sinks) of vortissance. For instance, a beam with just diagonal moments being different from zero will acquire real vortissance according to~Eqs.~(\ref{e_transp_solfringe_v}) to (\ref{e_transp_solfringe_vpp}) as
\begin{equation}
	\label{e_DV_fringe}
	V\,=\,2\kappa A \sqrt{\varepsilon_ 1^2 + \varepsilon _2^2 + 2\kappa ^2 A^2}\,.
\end{equation}
Imaginary vortissance can be created through skewed quadrupoles. The beam line introduced in section~\ref{s_LaVBeams} comprises a solenoid fringe field and (skewed) quadrupoles and it can form beams with real or with imaginary vortissance.

\section{periodic solutions of~$\mathcal{V}$ along pairs of solenoids}
Thanks to the analogy of transport of~$\mathcal{V}$ through solenoids and drifts to the transport of beam envelope through focusing quadrupoles and drifts, periodic solutions for both can be constructed for a given solenoid channel. Especially, for a channel made from solenoids with alternating field directions, the matrices transporting the envelope and~$\mathcal{V}$ are identical. Accordingly, the corresponding Twiss parameters are transported in the same way.

Starting from a given location~$s>s_0$ along the beam line, for beams with real vortissance~$V$, the Twiss parameters can be even equal (modulo the sign of the vorticity Twiss parameters), i.e.,
\begin{equation}
	\beta _v(s) \,=\, \pm\beta _e(s)
\end{equation}
\begin{equation}
	\alpha _v(s) \,=\, \pm\alpha _e(s)\,.
\end{equation}
Hence for real~$V$, permanent identity of envelope and vorticity Twiss parameters can be achieved.

This is not the case for imaginary~$V$. Although the vorticity Twiss parameters are transformed in the same way as the envelope Twiss parameters for any~$V$, their relation~$\beta _v \gamma _v - \alpha _v^2$=-1 forbids the vorticity Twiss parameters to be equal to the envelope Twiss parameters at any location of the beam line. Accordingly, these two sets of Twiss parameters are intrinsically and permanently different from each other. Periodic solutions for the Twiss parameters can be constructed for the envelope and for the vorticity. For real vortissance they are identical (modulo sign), while for imaginary vortissance they are intrinsically different.

\begin{figure}[hbt]
	\centering
	\includegraphics*[width=60mm,clip=]{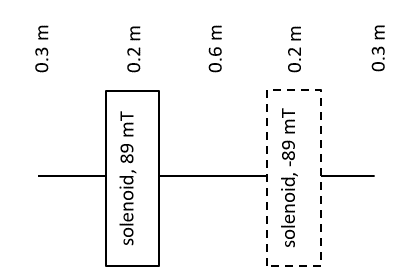}
	\caption{Pair of solenoids with opposite polarity.}
	\label{f_SolenoidPair}
\end{figure}

In the following, matched solutions shall be presented for the case of real~$V$ and of imaginary~$V$. The periodic channel comprises solenoids with effective lengths of 0.2~m being separated by 0.6~m from each other. Their magnetic field strength is 89~mT and its direction alternates from one solenoid to the next. Figure~\ref{f_SolenoidPair} depicts one solenoid pair comprising the smallest unit of the periodic channel.
The matrices of this unit are calculated from the individual transport matrices from subsection~\ref{ss_Vmatrices} to (in~m, 1, 1/m)
\begin{equation}
	m_u=
	\begin{bmatrix}
		0.695 &  1.42 &  0.00 & 0.00 \\
		-0.365 &  0.695 &  0.00 & 0.00 \\
		0.00 &  0.00 & 0.695 & 1.42 \\
		0.00 &  0.00 & -0.365 & 0.695
	\end{bmatrix}\,,
\end{equation}
\begin{equation}
	M_u=
	\begin{bmatrix}
		0.695 &  1.42 \\
		-0.365 &  0.695  \\
	\end{bmatrix}\,,
\end{equation}
and the unit's phase advance (horizontal, vertical, and vorticity) is accordingly~$|\Delta\Phi |$=46\degs. The periodic envelope Twiss parameters in both planes are~$\beta _e$=1.97~m and~$\alpha _e$=0.

Initially, the spinning beam is formed as described in section~\ref{s_LaVBeams}. Afterwards, four solenoids match the beam to the periodic channel.
Here, 2d-matching aims just for best envelope matching for each plane separately, i.e, full 4d-matching of all ten beam moments is not considered. Solenoids are used, since unlike quadrupoles, they strictly preserve angular momentum~$L$ and vortissance~$V$. The channel comprises eight solenoid pairs.

\begin{figure}[hbt]
	\centering
	\includegraphics*[width=88mm,clip=]{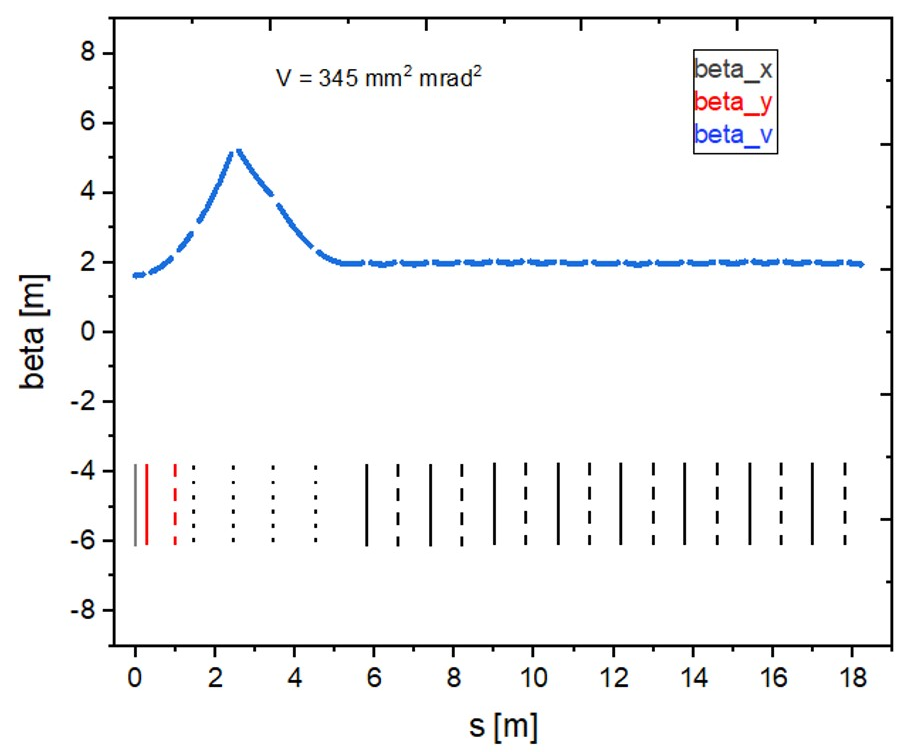}
	\caption{Beta functions along the full beam line comprising initial beam spinning, envelope matching to, and the periodic channel itself. Shown is the case of a beam with real vortissance, namely the $L$=$\mathcal{V}_A$-beam from section~\ref{s_LaVBeams}. Grey: horizontal, red: vertical, blue: vorticity~$\mathcal{V}$. The lines at the bottom indicate the beam line elements; from left to right: solenoid exit fringe field, regular quadrupole, skew quadrupole, four matching solenoids, eight solenoid pairs.}
	\label{f_envmatch_case10}
\end{figure}

Figure~\ref{f_envmatch_case10} plots the three~$\beta $-functions along the beam line for the~$L$=$\mathcal{V}_A$-beam shown in section~\ref{s_LaVBeams}. The vortissance is real, i.e., $V$=345~mm$^\text{2}$mrad$^\text{2}$. Matching to the periodic channel is perfect. The three~$\beta$-functions (horizontal, vertical, vorticity) are equal to each other all along the beam line. The periodicity is just the half of one solenoid pair. The Figure does not plot the $\beta$-functions inside of solenoids since they perform steep variations at the solenoids' entrance and exit. The latter cause non-symplectic transformations and just the complete solenoid comprises a symplectic element.

\begin{figure}[hbt]
	\centering
	\includegraphics*[width=88mm,clip=]{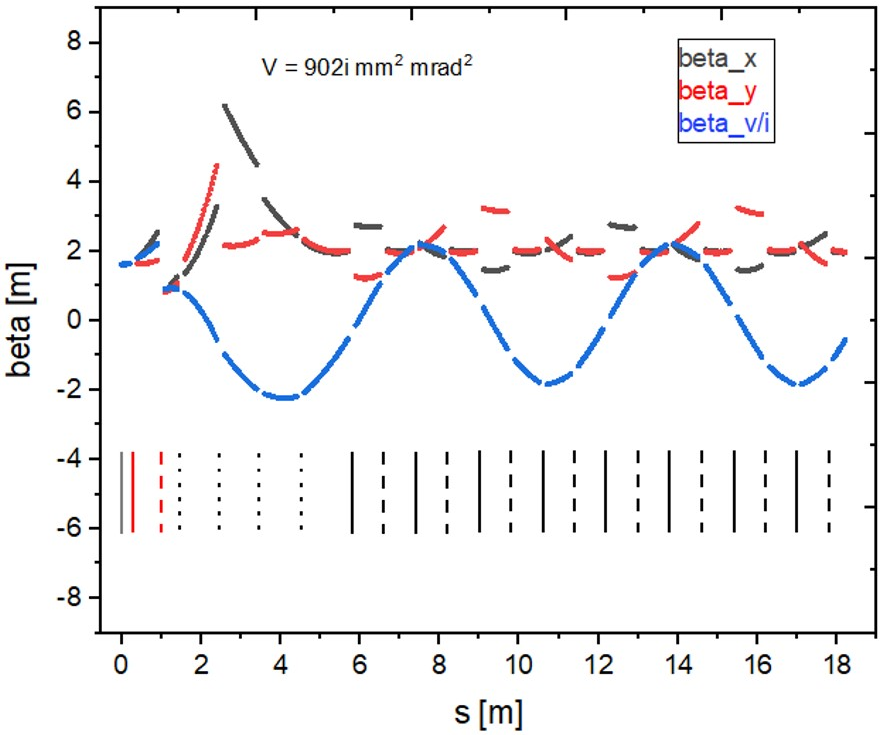}
	\caption{Beta functions along the full beam line comprising initial beam spinning, envelope matching to, and the periodic channel itself. Shown is the case of a beam with imaginary vortissance, namely the $\mathcal{V}_A$-beam from section~\ref{s_LaVBeams}. Grey: horizontal, red: vertical, blue: vorticity~$\mathcal{V}$. The lines at the bottom indicate the beam line elements; from left to right: solenoid exit fringe field, regular quadrupole, skew quadrupole, four matching solenoids, eight solenoid pairs.}
	\label{f_envmatch_case8}
\end{figure}

Instead, the~$\mathcal{V}_A$-beam with imaginary vortissance of~$V$=902i~mm$^\text{2}$mrad$^\text{2}$ behaves considerably different w.r.t.~the previous one. Figure~\ref{f_envmatch_case8} plots the respective $\beta$-functions along the beam line. This beam has no cylindrical symmetry and hence the two transverse~$\beta$-functions differ from each other. Additionally, as expected, the vorticity $\beta$-function differs from the transverse ones. Apart from being imaginary, also its amount is different from the one of the spatial~$\beta$-functions. The extension of quasi-periodicity of the three matched solutions is four solenoid pairs. It is called quasi-periodic, since the remaining vorticity mismatch parameter is still~0.05. Mismatch is defined analogue to the horizontal envelope mismatch parameter~\cite{wangler2}. For the vorticity~$\mathcal{V}$ this is due to the fact mentioned in subsection~\ref{ss_Vimaginary}: for imaginary~$V$, zero crossings of a periodic $\beta _v$-function may cause zero vorticity phase advance which is in contradiction to the non-zero trace of the respective matrix~$M_u$. Construction of a periodic vorticity $\beta _v$-function from less then four units would imply such crossings. Accordingly, such functions do not exist.

The vorticity~$\beta$-function performs a cosine-like oscillation. This is in full analogy to an oscillating envelope~$\beta$-function along a continuously focusing quadrupole channel. It confirms the finding that vorticity~$\mathcal{V}$ behaves along solenoid channels as beam envelope behaves along quadrupole channels.

\begin{figure}[hbt]
	\centering
	\includegraphics*[width=88mm,clip=]{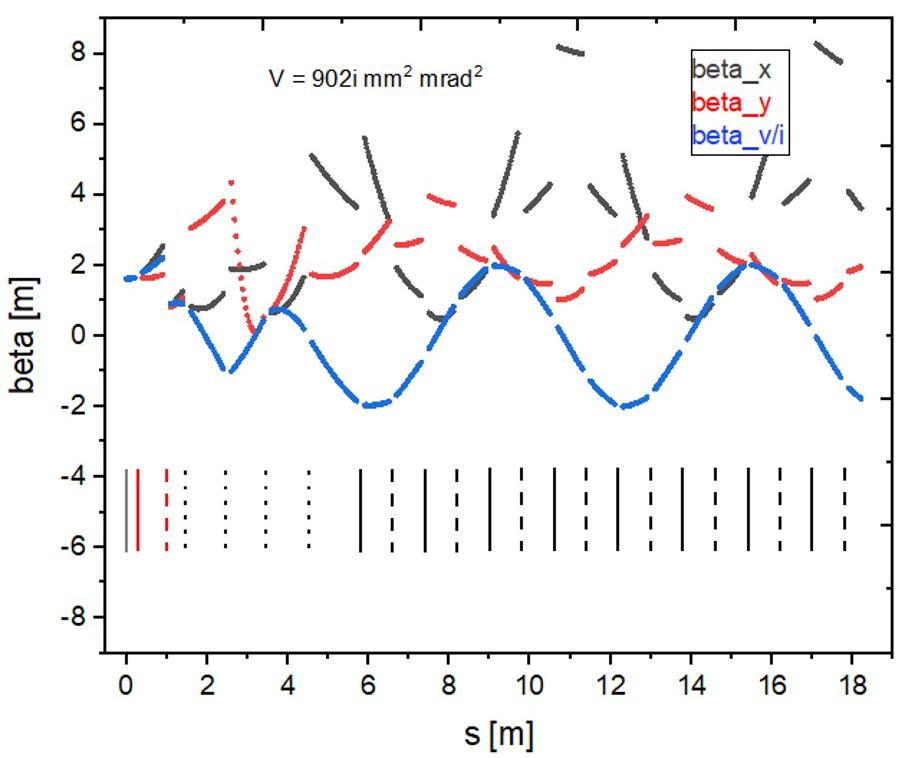}
	\caption{Beta functions along the full beam line comprising initial beam spinning, vorticity matching to, and the periodic channel itself. Shown is the case of a beam with imaginary vortissance, namely the $\mathcal{V}_A$-beam from section~\ref{s_LaVBeams}. Grey: horizontal, red: vertical, blue: vorticity~$\mathcal{V}$. The lines at the bottom indicate the beam line elements; from left to right: solenoid exit fringe field, regular quadrupole, skew quadrupole, four matching solenoids, eight solenoid pairs.}
	\label{f_WAmatch_case8}
\end{figure}

Finally, it shall be mentioned that matching to a periodic channel of solenoid pairs can be done either with the periodic envelope Twiss parameters or with the periodic vorticity Twiss parameters. Figures~\ref{f_envmatch_case10} and~\ref{f_envmatch_case8} showed examples for matching with the envelope Twiss parameters. In the case of a~$L$=$\mathcal{V}_A$-beam of Fig.~\ref{f_envmatch_case10} there is no difference. However, there is a difference for the $\mathcal{V}_A$-beam. Figure~\ref{f_WAmatch_case8} plots the vorticity matched counterpart of the case shown in Fig.~\ref{f_envmatch_case8}. Both cases result into periodic solutions but these solutions are different from each other.

\section{Pair of quadrupole triplets}
This section introduces a cell made from regular quadrupoles that is fully equivalent to a pair of solenoids w.r.t.~vorticity beam dynamics. The cell preserves the angular momentum and the vortissance. Its transport matrix reads as
\begin{equation}
	\label{e_m_singpart_TripletPair}
	m_{tp*}=
		\begin{bmatrix}
		a &  b &  0 & 0 \\
		c &  a &  0 & 0 \\
		0 &  0 &  a & b \\
		0 &  0 & c & a
	\end{bmatrix}\,,
\end{equation}
with $c=(a^2-1)/b$. This matrix is fully equivalent to the matrix of a pair of solenoids~$m_{sds}$ of Eq.~(\ref{e_m_singpart_sds}).

\begin{figure}[hbt]
	\centering
	\includegraphics*[width=88mm,clip=]{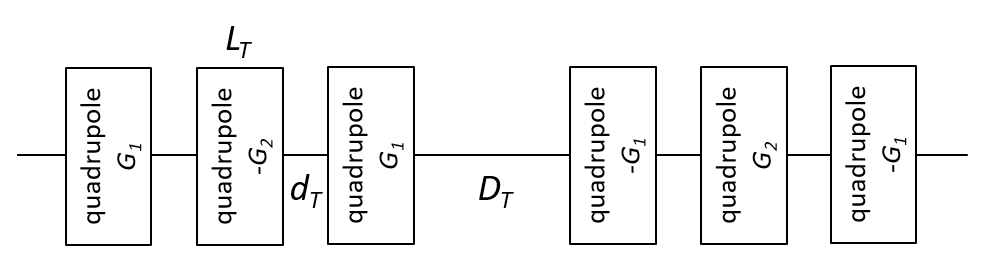}
	\caption{Pair of triplets from regular quadrupoles with opposite polarity. Each triplet comprises quadrupoles of same length~$L_T$. Seperation of the quadrupoles within a triplet is by a drift of length~$d_T$ and the triplets are separated by a drift of length~$D_T$. Outer quadrupoles have the gradient~$\pm G_1$ and center quadrupoles have the gradient~$\mp G_2$.}
	\label{f_TripletPair}
\end{figure}

The cell comprises two triplets of regular quadrupoles as sketched in~Fig.~\ref{f_TripletPair}.

The general transport matrix of such a pair of triplets is
\begin{equation}
	\label{e_m_singpart_Triplet}
	m_{tp}=
	\begin{bmatrix}
		d &  e &  0 & 0 \\
		g &  f &  0 & 0 \\
		0 &  0 &  f & e \\
		0 &  0 & g & d
	\end{bmatrix}\,,
\end{equation}
with $g=(df-1)/e$. This relation shall be stated here without presenting a stringent proof, which to the best of our knowledge is not at hand for the time being. It has been rather found with a computer code. Additionally, as expected, it has been found that by choosing an appropriate ratio~$G_2/G_1$, the triplet pair is set such, that Eq.~(\ref{e_m_singpart_TripletPair}) is met. Also this statement is not yet backed by an appropriate proof. However, for the time being, no example being in contradiction to one of the two statements has been found. Assuming that these statements are correct, all findings presented on vorticity beam dynamics along channels of solenoid pairs apply also to the dynamics along pairs of triplets.

\section{Conclusion and outlook}
Along channels from solenoids and properly set regular quadrupole triplets, the beam vorticity~$\mathcal{V}$ is modelled with remarkable similarity to the beam envelope. Transport matrices are identical and Twiss parameters are very similar. The vorticity counterpart to the rms emittance is the rms vortissance. However, the latter may be purely imaginary resulting into imaginary and negative Twiss parameters and even into zero current phase advances. Periodic vorticity can be achieved by lattices comprising pairs of solenoids and of appropriate quadrupole triplets.

Matching of~$L$=$\mathcal{V}_A$-beams with real vortissance is straight forward as for envelopes. Instead, matching of beams with imaginary vortissance ($\mathcal{V}_A$-beams) requires many more super-periods.

The presented studies are far from being concluded. Further activities shall aim on including the effects of space charge. The ultimate goal is application of the findings to reduction of emittance growth due to space charge through appropriate spinning and matching.

\section{Acknowledgement}
We thank Moses Chung (UNIST/Korea) and Chen Xiao (GSI/Germany) for valuable suggestions and advices during preparation of this report.

\appendix

\section{Definition of eigen-emittances and symplectic transformations}
\label{EigSymp}
Beam eigen-emittances are calculated through beam rms moments as
\begin{equation}
	\label{eigen12}
	\varepsilon_{1/2}\,=\,\frac{1}{2} \sqrt{-tr[(CJ)^2] \pm \sqrt{tr^2[(CJ)^2]-16\,det\,(C) }}\,,
\end{equation}
with
\begin{equation}
	\label{2nd_mom_matrix}
	C=
	\begin{bmatrix}
		\langle x^2 \rangle &  \langle xx'\rangle &  \langle xy\rangle & \langle xy'\rangle \\
		\langle xx'\rangle &  \langle x'^2\rangle & \langle yx'\rangle & \langle x'y'\rangle \\
		\langle xy\rangle &  \langle yx'\rangle &  \langle y^2\rangle & \langle yy'\rangle \\
		\langle xy'\rangle &  \langle x'y'\rangle & \langle yy'\rangle & \langle y'^2\rangle
	\end{bmatrix}\,,
\end{equation}
\begin{equation}
	\label{e_JMatrix}
	J=
	\begin{bmatrix}
		0 &  1 &  0 & 0 \\
		-1 &  0 &  0 & 0 \\
		0 &  0 &  0 & 1 \\
		0 &  0 & -1 & 0
	\end{bmatrix}\,,
\end{equation}
and $\varepsilon_{4d}\,$:=$\,(det\,C)^\frac{1}{2}\,$=$\,\varepsilon _1\cdot\varepsilon _2$. Projected transverse beam rms emittances are defined as~\cite{Floettmann_prstab}
\begin{flalign}
	\label{emrms_x}
	&\varepsilon _x \,=\,\sqrt{ \langle x^2 \rangle\langle x'^2\rangle\,-\,\langle xx' \rangle ^2}\,,\\
	\label{emrms_y}
	&\varepsilon _y \,=\,\sqrt{ \langle y^2 \rangle\langle y'^2\rangle\,-\,\langle yy' \rangle ^2}\,.
\end{flalign}
A matrix $M_s$ is called symplectic if it satisfies the equation
\begin{equation}
	M_s^TJM_s\,=\,J\,.
\end{equation}

\section{Relevant particle transport matrices and examples for transport of beam rms moments}
\label{s_PartTransp}
This appendix lists particle transport matrices being relevant for the manuscript. For some of them it gives examples for the corresponding transformation of some beam rms moments. In general, the full 4d-matrices are listed. In case that just the horizontal phase space is referred to, it is indicated by the subscript~$xx$. The beam rigidity is given by~$B\rho$ and the coordinate system defines~$x$ towards the left, $y$ upwards, and the beam direction~$s$ into forward direction, i.e., it is a right-handed system.

The matrix of a drift of length~$d$ is given by
\begin{equation}
	\label{e_m_singpart_drift}
	m_{drift}\,=\,
	\begin{bmatrix}
		1 & d & 0 & 0\\
		0 & 1 & 0 & 0 \\
		0 & 0 & 1 & d \\
		0 & 0 & 1 & 1
	\end{bmatrix}\,.
\end{equation}
For instance, the horizontal beam rms moments are changed by
\begin{equation}
	\label{e_transp_drift_xx}
	\langle x^2 \rangle \,\rightarrow\, \langle x^2 \rangle + 2d\langle xx' \rangle +d^2 \langle x'^2 \rangle \,,
\end{equation}
\begin{equation}
	\label{e_transp_drift_xxp}
	\langle xx' \rangle \,\rightarrow\, \langle xx' \rangle +d\langle x'^2 \rangle \,,
\end{equation}
\begin{equation}
	\label{e_transp_drift_xpxp}
	\langle x'^2 \rangle \,\rightarrow\, \langle x'^2 \rangle \,.
\end{equation}

A horizontally focusing quadrupole with effective length~$l$ and magnetic field gradient~$G$ is modelled by
\begin{equation}
	\label{e_m_singpart_quad}
	m_{q,xx}\,=\,
	\begin{bmatrix}
		cos\,\Omega & u^{-1}\,sin\,\Omega \\
		-u\,sin\,\Omega & cos\,\Omega
	\end{bmatrix}\,,
\end{equation}
with $\Omega$:=$ul$ and $u$:=$|G/(B\rho)|^\frac{1}{2}$. If the quadrupole is short ($ul\ll 1$), this matrix is well approximated by
\begin{equation}
	\label{e_m_singpart_shortquad}
	m_{sq,xx}\,=\,
	\begin{bmatrix}
		1 & 0 \\
		-k & 1
	\end{bmatrix}\,,
\end{equation}
where~$k:$=$ul$. The short quadrupole transforms the horizontal moments as
\begin{equation}
	\label{e_transp_shortquad_xx}
	\langle x^2 \rangle \,\rightarrow\, \langle x^2 \rangle \,,
\end{equation}
\begin{equation}
	\label{e_transp_shortquad_xxp}
	\langle xx' \rangle \,\rightarrow\, \langle xx' \rangle -k \langle x^2 \rangle \,,
\end{equation}
\begin{equation}
	\label{e_transp_shortquad_xpxp}
	\langle x'^2 \rangle \,\rightarrow\, \langle x'^2 \rangle -2k \langle xx' \rangle +k^2 \langle x^2 \rangle \,.
\end{equation}

The skewed quadrupole is a quadrupole being rotated by 45\degs ~clockwise around the positive beam direction. For a short skew quadrupole the matrix reads as
\begin{equation}
	\label{e_m_singpart_skewquad}
	m_{q,skew}\,=\,
	\begin{bmatrix}
		1 & 0 & 0 & 0\\
		0 & 1 & -k & 0 \\
		0 & 0 & 1 & 1 \\
		-k & 0 & 0 & 1
	\end{bmatrix}\,.
\end{equation} 

A solenoid of length~$L_s$ shall have the longitudinal magnetic field strength~$B$ along the positive beam direction. The solenoid's strength is defined by 
$\kappa\,:=\frac{B}{2(B\rho)}$. Additionally, $C\,:=\,cos(\kappa L_s)$, and $S\,:=sin(\kappa L_s)$. The solenoid transport matrix is
\begin{equation}
	\label{e_m_singpart_sol}
	m_{sol}\,=\,
	\begin{bmatrix}
		C^2 &  \frac{CS}{\kappa} &  CS & \frac{S^2}{\kappa} \\
		-\kappa CS &  C^2 & -\kappa S^2 & CS \\
		-CS &  -\frac{S^2}{\kappa} &  C^2 & \frac{CS}{\kappa} \\
		\kappa S^2 & -CS & -\kappa CS & C^2
	\end{bmatrix}\,.
\end{equation}
The following example gives the lengthy transformation of the beam moment~$\langle xy' \rangle $ by a solenoid
\begin{equation}
	\label{e_transp_sol_xyp}
	\begin{split}
		\langle xy' \rangle \,\rightarrow\,& +\langle x^2 \rangle\cdot \kappa S^2C^2 +\langle xx' \rangle\cdot \left[ CS^3-C^3S \right] \\
		&+\langle xy \rangle\cdot \left[ \kappa CS^3-\kappa C^3S \right]  +\langle xy' \rangle\cdot \left[ C^4+S^4 \right] \\
		&-\langle x'^2 \rangle\cdot C^2S^2/\kappa  -\langle x'y \rangle\cdot 2C^2S^2 \\
		&+\langle x'y' \rangle\cdot \left[ C^3S/\kappa - CS^3/\kappa \right]  -\langle y^2 \rangle\cdot \kappa C^2S^2 \\
		&+\langle yy' \rangle\cdot \left[ C^3S-CS^3 \right]  +\langle y'^2 \rangle\cdot \left[ C^2S^2/\kappa \right]\,.
	\end{split}
\end{equation}
For a short solenoid holds $\kappa L_s\,\ll 1$ which results into
\begin{equation}
	\label{e_m_singpart_solshort}
	m_{sol,short}\,=\,
	\begin{bmatrix}
		1 & 0 &  0 & 0 \\
		-\kappa ^2L_s &  1 & 0 & 0 \\
		0 & 0 & 1 & 0 \\
		0 & 0 & -\kappa ^2 L_s & 1
	\end{bmatrix}\,.
\end{equation}

The solenoid's entrance and exit field are part of the solenoid matrix. However, they can be modelled stand-alone through:
\begin{equation}
	\label{e_m_singpart_solfringe}
	m_{sol,fringe}\,=\,
	\begin{bmatrix}
		1 & 0 & 0 & 0 \\
		0 & 1 & \pm \kappa & 0 \\
		0 & 0 & 1 & 0 \\
		\mp \kappa & 0 & 0 & 1
	\end{bmatrix}\,,
\end{equation}
where the upper (lower) sign refers to the entrance (exit) fringe field. This fringe field matrix is the only one being mentioned within this study which is not symplectic. It changes the two beam eigen-emittances. However, it preserves the 4d-rms emittance which is the product of both.

Of special relevance here is the sequence of a solenoid, a drift, and a second solenoid with opposite field w.r.t.~the direction of the first one. The corresponding transport matrix does not couple the two transverse planes and reads as
\begin{equation}
	m_{sds}\,:=\,m_{sol}(-B)\cdot m_{drift}\cdot m_{sol}(B)\,,
\end{equation}
\begin{equation}
	\label{e_m_singpart_sds}
	m_{sds}\,=\,
	\begin{bmatrix}
		m_{sds,xx} & 0 \\
		0 &  m_{sds,yy}
	\end{bmatrix}\,,
\end{equation}
\begin{equation}
	\begin{split}
		&m_{sds,xx}\,=\,m_{sds,yy}\,= \\
		&\begin{bmatrix}
			C^2 - \kappa dCS - S^2 &  2\frac{CS}{\kappa} + dC^2 \\
			-2\kappa CS + \kappa ^2 dS^2 &  C^2 - \kappa dCS - S^2
		\end{bmatrix}\,.
	\end{split}
\end{equation}

\section{Transport of transverse beam envelope}
\label{s_EnvTransp}
In the following, the discussion is restricted to the horizontal plane, however, the vertical one could have been chosen as well. The beam moments are expressed through the derivatives of the squared beam size~$\langle x^2 \rangle $ w.r.t.~$s$ along a drift:
\begin{equation}
	\langle x^2 \rangle '\,=\,2\langle xx'\rangle \,,
\end{equation}
\begin{equation}
	\langle x^2 \rangle ''\,=\,2\langle x'^2\rangle	\,,
\end{equation}
and explicitly mentioning
\begin{equation}
	\langle x^2 \rangle '''\,=\,0	\,.
\end{equation}
Defining the envelope vector~$\vec{X}:=[\langle x^2 \rangle \,,\langle x^2 \rangle '\,, \langle x^2 \rangle '']$, the transport of the beam envelope along a drift is expressed through a matrix equation
\begin{equation}
	\label{e_m_drift_e_3}
\vec{X}\,\rightarrow\,
	\begin{bmatrix}
		1 & d & \frac{d^2}{2} \\
		0 & 1 & d \\
		0 & 0 & 1
	\end{bmatrix}\cdot
\vec{X}\,,
\end{equation}
which is just the Taylor series.

In an analogue way, the transport of the envelope through a short quadrupole is stated as
\begin{equation}
	\label{e_m_shortquad_e_3}
\vec{X}\,\rightarrow\,
	\begin{bmatrix}
		1 & 0 & 0 \\
		-2k & 1 & 0 \\
		2k^2 & -2k & 1
	\end{bmatrix}\cdot
\vec{X}\,.
\end{equation}

Finally, the transport of the envelope through a general horizontally focusing quadrupole is
\begin{equation}
	\label{e_m_quad_e_3}
	\vec{X}\,\rightarrow\,
	\begin{bmatrix}
		C_q^2 & \frac{C_qS_q}{u} & \frac{S_q^2}{2u^2}\\
		-2uC_qS_q & C_q^2-S_q^2 & \frac{C_qS_q}{u} \\
		2u^2S_q^2 & -2uC_qS_q & C_q^2
	\end{bmatrix}\cdot
	\vec{X}\,,
\end{equation}
where $C_q$ and $S_q$ refer to the trigonometrical functions stated in~Eq.~(\ref{e_m_singpart_quad}).


\end{document}